\documentclass[sts]{imsart}

\RequirePackage{amsthm,amsmath,amsfonts,amssymb, bm}
\RequirePackage[authoryear]{natbib}
\RequirePackage{graphicx}
\RequirePackage{xcolor}
\RequirePackage{ulem}
\RequirePackage{hyperref}

\startlocaldefs

\usepackage{sidecap}
\newcommand{\bfg}{\textbf{g}}

\endlocaldefs

\begin{document}

\begin{frontmatter}


\title{Statistical Challenges in Tracking the Evolution of SARS-CoV-2}
\runtitle{Statistical challenges in phylodynamics}

\begin{aug}
\author[A]{\fnms{Lorenzo} \snm{Cappello}\ead[label=e1]{cappello@stanford.edu}},
\author[B]{\fnms{Jaehee} \snm{Kim}\ead[label=e2]{jk2236@stanford.edu}},
\author[C]{\fnms{Sifan} \snm{Liu}\ead[label=e3]{sfliu@stanford.edu}}
\and
\author[D]{\fnms{Julia A.} \snm{Palacios}\ead[label=e4]{juliapr@stanford.edu}}
\address[A]{Lorenzo Cappello is a Postdoctoral Research Fellow in the Department of Statistics, Stanford University, Stanford, CA 94305, USA \printead{e1}.}
\address[B]{Jaehee Kim is a Postdoctoral Research Fellow in the Department of Biology, Stanford University, Stanford, CA 94305, USA \printead{e2}.}
\address[C]{Sifan Liu is a Ph.D. student in the Department of Statistics, Stanford University, Stanford, CA 94305, USA \printead{e3}.}
\address[D]{Julia A. Palacios is an Assistant Professor in the Departments of Statistics and Biomedical Data Sciences, Stanford University, Stanford, CA 94305, USA. \printead{e4}.}
\end{aug}

\begin{abstract}
Genomic surveillance of SARS-CoV-2 has been instrumental in tracking the spread and evolution of the virus during the pandemic. The availability of SARS-CoV-2 molecular sequences isolated from infected individuals, coupled with phylodynamic methods, have provided insights into the origin of the virus, its evolutionary rate, the timing of introductions, the patterns of transmission, and the rise of novel variants that have spread through populations. Despite enormous global efforts of governments, laboratories, and researchers to collect and sequence molecular data, many challenges remain in analyzing and interpreting the data collected. Here, we describe the models and methods currently used to monitor the spread of SARS-CoV-2, discuss long-standing and new statistical challenges, and propose a method for tracking the rise of novel variants during the epidemic.
\end{abstract}




\begin{keyword}
\kwd{Phylodynamics}
\kwd{genetic epidemiology}
\kwd{coalescent}
\kwd{Bayesian nonparametrics}
\kwd{birth-death processes}
\kwd{SIR models}
\end{keyword}

\end{frontmatter}

\section{Introduction}
In the last couple of years, we have witnessed an unprecedented global effort to collect and share SARS-CoV-2 molecular data and sequences. This effort has resulted in more than two million molecular sequences being available for download in public repositories such as GISAID \citep{shu2017gisaid} and GenBank today. These viral RNA sequences are \textbf{consensus}\footnote{A glossary in the appendix explains the terms in bold that not all statisticians may be familiar with.} sequences of about 30,000 nucleotides isolated from biological samples, such as nasal swabs, from infected individuals. Analyses of viral molecular sequences provide evidence of human-to-human transmission and allow the investigations of SARS-CoV-2 origins \citep{andersen2020proximal,boni2020evolutionary}. Moreover, they are routinely used to investigate outbreaks \citep{maccannell2021genomic,deng2020genomic}, track the speed and spread of viral transmission across the world \citep{nextstrain}, and monitor the evolution of new variants \citep{vol21}. 

The field of phylodynamics of infectious diseases, also referred to as molecular epidemiology, aims to understand disease dynamics by joint modeling of evolutionary, immunological, and epidemiological processes \citep{Grenfel2004,volz2013viral}. It is assumed that these processes shape the underlying  viral phylogeny of a sample of molecular sequences at a \textbf{locus}. Under models of neutral evolution, it is assumed that a process of \textbf{substitutions} is superimposed along the branches of the phylogeny, generating the observed variation in the sample of molecular sequences. More complex evolutionary models consider the effects of other types of \textbf{mutations} and sources of variation, such as \textbf{recombination} and \textbf{selection} \citep{wak09}.

\begin{figure}[!htbp]
	\centering
	\includegraphics[width=0.9\textwidth]{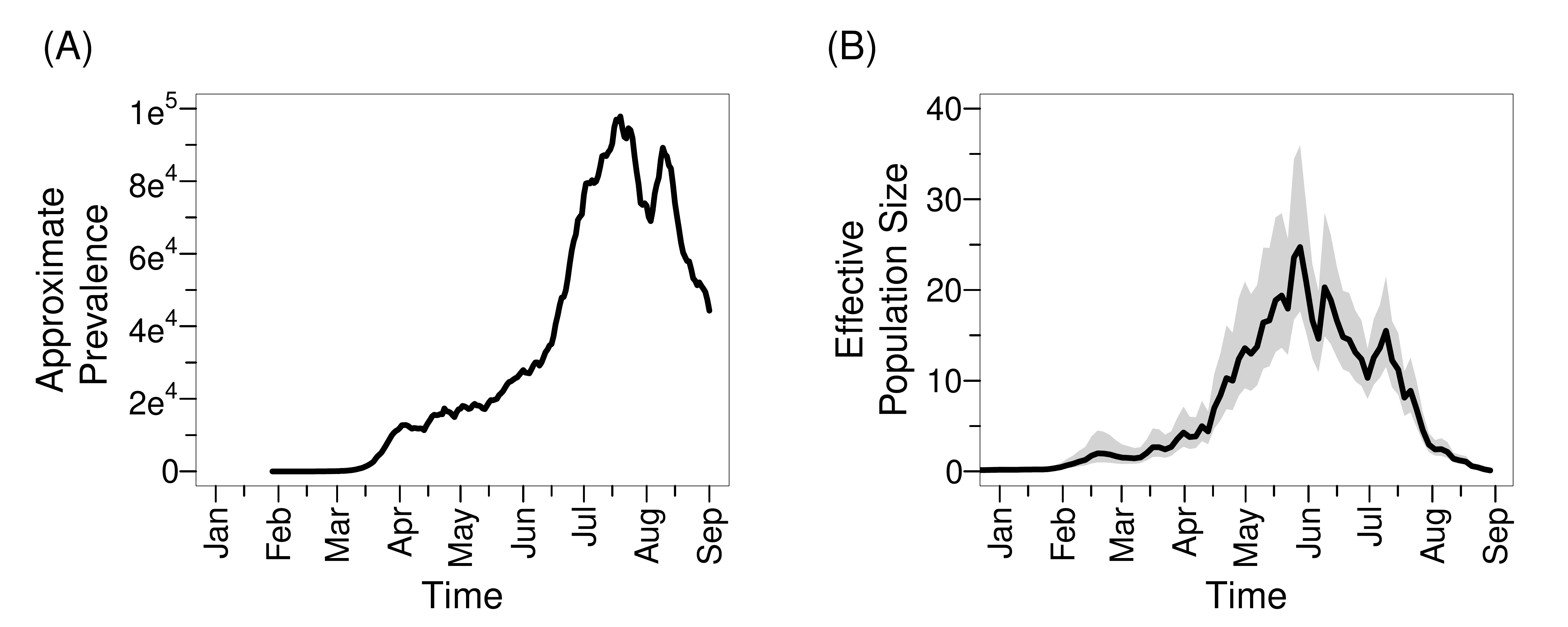}
	\caption{\small{\textbf{Phylodynamic analysis of SARS-CoV-2 sequences in California in 2020.} (A) 10-days cumulative sum of the daily number of new cases in California. (B) Posterior median of the effective population size (black line) and $95\%$ credible region (gray area). Model and data details appear in Appendix.}}
	\label{fig:cali}
\end{figure}

A viral phylogeny is a timed bifurcating tree that represents the ancestral history of a sample at a locus (Figure \ref{fig:het_tree}(A)). This viral phylogeny can be obtained by maximum parsimony methods or by maximum likelihood from observed molecular sequences \citep{felsenstein2004inferring}. In the case of maximum likelihood, a model of substitutions (or mutations) is required. In phylodynamics, however, the study usually does not end at a single phylogeny. The aim is to understand the evolutionary and epidemiological forces that shape the phylogeny. To this end, 
the phylogeny is typically assumed to be the realization of either a birth–death-sampling process (BDSP) \citep{Stadler2013} or a coalescent process (CP) \citep{king82,rodrigo99b}.
In the context of disease dynamics, the BDSP is parameterized by the 
transmission rate $\lambda(t)$ and recovery rate $\mu(t)$, all of which are parameters of interest in epidemiology and public health. The CP is parameterized by the effective population size $N_{e}(t)$, a measure of relative genetic diversity over time that serves as a proxy of the growth and decline in the number of infections over time. For example, Figure~\ref{fig:cali}(B) shows the estimated effective population size of SARS-CoV-2 in California in the first nine months of 2020, together with panel (A) that shows the 10-day cumulative number of new cases.

It is possible to link epidemiological compartmental models, such as the susceptible-infected-recovered (SIR) model, to phylogenies via the CP \citep{Volz2014, Boskova2014}. With the simplest SIR model, the coalescent effective population size $N_{e}(t)$ is expressed in terms of the number of infections over time, transmission rate, and the number of susceptible individuals. More complex population dynamics and compartmental models can also be incorporated into the CP framework \citep{Volz2018}. We refer to the general class of such models as the CP-EPI. In this paper, we survey current methods and challenges for estimating epidemiological parameters from the BDSP and the CP-EPI frameworks and their applications in studying the evolution and epidemic spread of SARS-CoV-2.

Although the field of phylodynamics has advanced in recent years, it has been recognized that there are still many challenges in using sequence data to infer disease dynamics. In \citet{frost2015eight}, the authors stated the following challenges: (1) modeling of more complex evolutionary processes such as recombination, selection, within-host evolution, population structure, and stochastic population dynamics; (2) modeling of more complex sampling scenarios, (3) joint modeling of phenotypic and genetic data, and (4) computation. We have subsequently seen advances in solving some of these challenges, such as modeling of recombination \citep{muller2021recombination} and stochastic population dynamics \citep{Stadler2013a, Volz2018}, incorporation of more complex sampling scenarios \citep{kar16,kar20b,par20,cap21}, and joint modeling of epidemiological and genetic data \citep{Li2017, Tang2019, Zarebski2021, Featherstone2021}. However, even in the simplest evolutionary model, inference involves integration over the high dimensional space of phylogenies. This is usually achieved via Markov chain Monte Carlo (MCMC) methods, making inference computationally intractable for large sample sizes. 

Apart from the existing challenges, the pandemic presented us with new statistical challenges. Here, we focus our discussion on four challenges: (1) scalability, (2) phylodynamic hypotheses testing, (3) adaptive modeling of the sampling process and (4) interpretability of model parameters. 

Current phylodynamic implementations are computationally incapable of analyzing the amount of SARS-CoV-2 sequences available; researchers are forced to subsample available data and to sacrifice model complexity. In Section~\ref{sec:scalability}, we focus on the scalability of Bayesian phylodynamic methods. We provide an overview of current practices for analyzing SARS-CoV-2, recent advances in Bayesian computation and the particular challenges in applying such advances in phylodynamics. 

The continual rise of new SARS-CoV-2 variants with putative higher transmissibility, demands for novel strategies for statistical hypotheses tests that not only rely on molecular data but also on the sampling process of sequences and phenotypic information from the host and the pathogen. In Section~\ref{sec:test}, we provide an overview of current practices for testing higher transmissibility of variants of concern and provide a new semi-parametric model that allows for this testing. 

Heterogeneous strategies of molecular sequence collection demands for adaptive phylodynamic methods that properly account for this heterogeneity. In Section~\ref{sec:pref}, we discuss recent advances in temporal modeling of the sampling process of molecular sequences. Finally, increasing the interpretability of model parameters is becoming one of the most important challenges in phylodynamic inference. Meaningful parameterization often requires more complex modeling and inferential challenges. In Section~\ref{sec:sir}, we provide an overview of phylodynamic methods that aim to infer prevalence and other epidemiological parameters from molecular sequences and count data, and highlight some future directions in the field. Section~\ref{sec:disc} concludes with a discussion of encompassing themes that have emerged in the paper.

\section{Background}

Neutral models of evolution typically assume that the tree topology and the branching (or coalescent times) are independent. In the next two sections, we will summarize the two most popular models on phylogenies used in phylodynamics.



\subsection {Coalescent process (CP).} A retrospective probability model on phylogenies is the standard coalescent. The standard coalescent was initially proposed as the limiting stochastic process of the ancestry of $n$ samples chosen uniformly from a large population of $N\gg n$ individuals undergoing simple forward dynamics \citep{king82,kingn82}. It was later extended to variable population sizes \citep{slatkin_pairwise_1991,gri94b} and heterochronous sampling \citep{rod99}. Here, we consider these extensions, and assume that samples are obtained at times $\mathbf{y}=(y_{1},\ldots,y_{n})$, with $y_{i}$ denoting the sampling time of the $i$th sample. Coalescent models have been reviewed extensively \citep{rosenberg2002genealogical,marjoram2006modern,tav04,wak09,berestycki2009recent,wakeley2020developments} and we refer the reader to those references for further details. 

\begin{figure}[!htbp]
	\centering
	\includegraphics[scale=0.6]{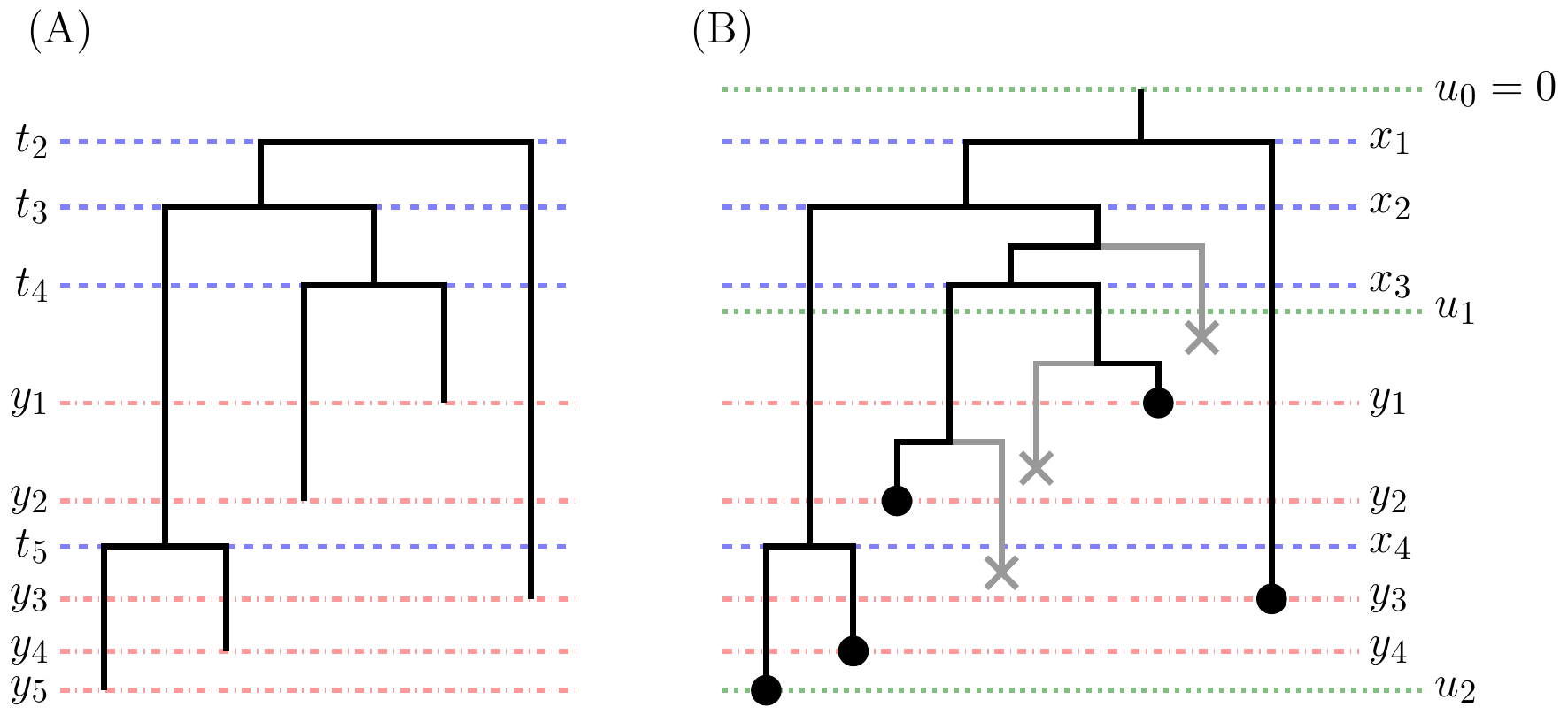}
	\caption{\small{\textbf{Example of a phylogeny.} (A) Example of a phylogeny realization from the CP with $n=5$. $t_i$'s and $y_i$'s indicate coalescent times and sampling times, respectively. (B) An example phylogeny from the BDSP that started at $t=u_0$ and ended at $u_2$. The filled circles represent sampled lineages and the crosses indicate extinct lineages. At each time interval $[u_{i-1}, u_i)$, the rate parameters are assumed to be constant. The branching times and the serial sampling times are denoted by $x_k$ and $y_k$, respectively. Lineages can also be sampled contemporaneously at each $u_i$ with probability $\rho_i$.}}
	\label{fig:het_tree}
\end{figure}

The space of phylogenies is the product space $\mathcal{G}_{n}=\mathcal{T}_{n} \times \mathbb{R}^{n-1}$ of discrete ranked and labeled tree topologies $\mathcal{T}_{n}$ and of vectors of coalescent times $\mathbf{t}=(t_{2},\ldots,t_{n})$, where $t_{k}$ indicates the $(n-k+1)$-th time two lineages have a common ancestor, when proceeding backwards in time from the tips to the root (Figure \ref{fig:het_tree}(A)). The coalescent density of the phylogeny is:
\begin{align} \label{eq:coal}
    p(\mathbf{g}\mid N_{e}(t))=\exp\left(-\int^{\infty}_{0}\frac{C(t)}{N_{e}(t)}dt \right)\prod^{n}_{k=2}\frac{1}{N_{e}(t_{k})},
\end{align}
where $C(t)=\frac{A(t)(A(t)-1)}{2}$, termed the coalescent factor, is a combinatorial factor of the number of extant lineages $A(t)=\sum^{n}_{i=1}I(y_{i}>t)-\sum^{n}_{k=2}I(t_{k}>t)$. Here, the density is parameterized by $(N_e(t))_{t\geq 0}:=N_e$ that denotes the effective population size (EPS). In the CP, the rate of coalescence, which is when two lineages meet a common ancestor, is inversely proportional to the EPS. That is, going backwards in time, a long waiting time for the first coalescence indicates large EPS during that period of time. Under Wright-Fisher population dynamics, $N_{e}(t)=N(t)/N(0)$ is the relative census population size \citep{tav04}. Under more general population dynamics, the EPS is usually interpreted as a relative measure of genetic diversity as it might not depend linearly on the census population size \citep{wakeley2009extensions}. 


\subsection{Birth-death-sampling process (BDSP)} \label{subsed:intro.bd}
In the BDSP \citep{Stadler2013a}, the population dynamics follows an inhomogeneous birth-death Markov process forward in time in which a birth represents a transmission event, and a death represents the event in which the individual either recovers, becoming noninfectious, or dies. The process starts with a single infected individual at time $t=0$. At time $t$, a transmission occurs with rate $\lambda(t)$ and an individual becomes noninfectious with rate $\mu(t)$. Given that we observed a fraction of the population, BDSP requires the definition of a sampling process. The sampling process selects either single lineages according to a Poisson process with rate $\psi(t)$, or in bulk at predetermined fixed time points with a sampling probability $\rho(t)$ of each lineage. Figure~\ref{fig:het_tree}(B) depicts a full realization of the process in which only black tips are sampled to form the sampled phylogeny. 

Current implementations of the BDSP \citep{Stadler2013a,bou19} assume all rates are piecewise constant functions with jumps at $u_1< \dots< u_{p-1}$ and denoted in vector form by $\bm{\lambda}, \bm{\mu}, \bm{\psi},$ and $\bm{\rho}$, where $i$-th element is a rate in $[u_{i-1}, u_i)$ ($i=1, \dots, p$). $s$ tips are sequentially sampled at $y_1 < \dots < y_s$, and additionally, $m_i$ lineages are sampled in bulk at each time $u_i$ with the sampling probability $\rho_i$ of each lineage, resulting in $n = s + \sum_{i=1}^{p} m_i$ total samples. The $n-1$ branching times of the $n$ samples are denoted by $x_1 < \dots < x_{n-1}$, and let $n_i$ be the number of lineages in the phylogeny at time $u_i$ excluding newly sampled lineages at $u_i$. 
The BDSP phylogeny density is 
%
\begin{align} \label{eq:bd.prob}
	p(\mathbf{g} \mid \bm{\lambda}, \bm{\mu}, \bm{\psi}, \bm{\rho}, \bm{t}) &= \underbrace{q_1(0)}_{\small{\text{trans. at the root}}} \underbrace{\prod_{i=1}^{n-1} \lambda_{I(x_i)} q_{I(x_i)}(x_i)}_{\small{\text{trans. at internal nodes}}} \underbrace{\prod_{i=1}^s \frac{\psi_{I(y_i)}}{q_{I(y_i)}(y_i)}}_{\small{\text{seq. sampling trans.}}} \nonumber \\
	&\times \underbrace{\prod_{i=1}^p \left( \frac{\rho_i}{q_{i}(u_{i})}\right)^{m_i}}_{\small{\text{bulk. sampling trans.}}}\underbrace{\prod^{p-1}_{i=1} \left( \frac{(1-\rho_i) q_{i+1}(u_i)}{q_{i}(u_{i})} \right)^{n_i}}_{\small{\text{no trans. among $n_{i}$ extant lineages}}},
\end{align}
where $I(t)=i$ ($i=1, \dots, p$) for $t \in [u_{i-1}, u_i)$ and 0 otherwise. $q_{i}(t)$ denotes the density of the per-lineage dwelling time in $[u_{i-1},u_{i})$, that is, the density that a lineage at time $t \in [u_{i-1},u_{i})$ evolves as observed in the tree, with $q_i(u_i)=1$. The explicit expression for $q_{i}(t)$ appears in the Supplementary Material of \citet{Stadler2013a}. 

Equation~\eqref{eq:bd.prob} is the result of a series of papers. 
\citet{Thompson1975} showed that in the case of constant birth and death rates, the branching times of the tree with tips consisting of only present-day individuals, conditioned on the time at the root, are i.i.d. \citet{nee1994reconstructed} and \citet{gernhard2008conditioned} showed that the same result can be obtained when conditioning on the number of tips. The fact that Equation~\ref{eq:bd.prob} can be obtained as a completely observed Markovian process is the result of \citet{stadler2009incomplete, Stadler2010}, who showed that the BDSP can be interpreted as a birth-death process with reduced rates and complete sampling. Finally, \citet{Stadler2013a} extended the result under piece-wise constant birth, death and sampling rates. Here, branching times are not longer i.i.d. but remain independent. 

Extensions to the BDSP include the flexibility of modeling the probability $r(t)$ of a sampled lineage to become effectively noninfectious immediately following the sampling event, and the modeling of multi-type birth and death events, accounting for population structure \citep{scire2020improved}. A more general framework unifying existing BDSP models has been recently proposed by \citet{MacPherson2021}.

\subsection{Bayesian Phylodynamic Inference}
Phylogenies are usually not observed; the CP or the BDSP density is used as prior on the phylogeny in order to infer phylodynamic parameters denoted by $\boldsymbol{\theta}$ such as $N_{e}(t)$ or transmission rate $\lambda(t)$. Let $D$ denote the observed molecular sequences sampled at times $\mathbf{y}$. In the phylodynamic generative model, phylodynamic parameters stochastically dictate the shape of the phylogeny; given a phylogeny, a process of substitutions is superimposed along the branches of the phylogeny that generates observed data. The target posterior distribution is the augmented posterior $P(\mathbf{g},\boldsymbol{\theta}, \mathbf{Q} \mid D,\mathbf{y})$, where $\mathbf{Q}$ denotes substitution parameters. \citet{yang2014molecular} provides a comprehensive reference of different mutation models used in phylodynamics. 

\section{Scalability} \label{sec:scalability}



The posterior distribution $P(\mathbf{g},\boldsymbol{\theta}, \mathbf{Q} \mid D,\mathbf{y})$ is usually approximated via Markov chain Monte Carlo (MCMC). Mixing of Markov chains in the high dimensional space of phylogenetic trees and model parameters is challenging, mostly because the posterior distributions on these discrete-continuous state spaces are highly multimodal \citep{whi15}. State-of-the-art algorithms, such as those implemented in BEAST \citep{suchard2018bayesian} and BEAST2 \citep{bou19}, exploit GPUs \citep{ayres2012beagle} and multi-core CPUs to run multiple MCMC chains in parallel, and carefully designed transition kernels to improve the mixing.

A parallel tempering method proposed by \cite{altekar2004parallel} apply the Metropolis-coupled MCMC ($\text{MC}^3$) method in which multiple chains are run in parallel and "heated". Here, the posterior term in the acceptance ratio is raised to a power (temperature). After a certain number of iterations, two chains are selected to swap states, encouraging them to explore the parameter space and prevent them from getting stuck in a peak.  \cite{muller2020adaptive} improve upon this $\text{MC}^3$ method by choosing the temperatures adaptively.

Despite these efforts, current methods can only be applied to hundreds or few thousands of samples and thus have limited applicability to pandemic-size datasets. The main bottleneck in these algorithms is the exploration of the space of phylogenetic trees. Under the substitution models typically used for phylodynamic inference, all phylogenies with $n$ tips have non-zero likelihood, and Markov chains on the space of phylogenetic tree topologies are known to mix in polynomial time \citep{simper2020adjacent}.
In the rest of this section, we first summarize some of the most popular pipelines recently used for phylodynamic analyses of SARS-CoV-2 sequences. Then we review some of the recent advances towards scalable phylodynamic inference.

\subsection{Practices in analyzing SARS-CoV-2 data}
Lacking a method or software capable of dealing with the number of available sequences, researchers usually resort to different types of approximations: (1) partition available data into subsets and analyze each subset independently \citep{lemey2020accommodating,vol21}, or (2) analyze a subsample selected at random from the set of available sequences \citep{choi2020phylodynamic,muller2021viral}, or (3) estimate a single MLE phylogeny from subsampled sequences, for example, the phylogeny available and periodically updated in Nextstrain \citep{nextstrain}, or obtain an MLE phylogeny directly with fast implementations such as TreeTime \citep{sagulenko2020treetime} and IQ-TREE \citep{minh2020iq}; phylodynamic parameters are then inferred from the fixed phylogeny  \citep{van2020no,maurano2020sequencing,dellicour2021phylodynamic}. 

Examples of the largest scale analyses have been  \cite{vol21}, who include approximately $27,000$ sequences and \cite{du2021establishment}, who study 50,887 SARS-CoV-2 genomes in the UK. \cite{du2021establishment} divide the full dataset into five smaller datasets according to whether the samples carry one of five groups of mutations. These five groups of mutations partition the dataset into five different lineages. The authors then estimate the five phylogenies with an approximate MLE method, where they employ an approximate likelihood in lieu of an exact one. The five MLE phylogenies are then analyzed separately. Phylogenies obtained with MLE methods cannot be readily used to infer evolutionary parameters in a CP framework if they are multifurcating trees. This is generally the case in SARS-CoV-2 applications. To infer the EPS, the authors sample over the set of binary trees compatible with a given multifurcating tree. 
Here, a binary phylogeny is compatible with a multifurcating phylogeny if the latter can be obtained by removing internal nodes from the binary phylogeny. 
Let $\mathbf{g}_{MLE}$ denote the estimated MLE phylogeny. 
The authors then approximate the posterior distribution $P(\mathbf{g} \prec \mathbf{g}_{MLE}, N_{e}(t), \mathbf{Q} \mid D,\mathbf{g}_{MLE})$ while constraining the posterior exploration to binary phylogenies that are compatible with the MLE phylogeny. Although this method does not account for all phylogenetic uncertainty, it does offer a solution to deal with multifurcating trees.
 \cite{vol21} first estimate the MLE phylogeny, 
 then identify on the MLE phylogeny several clades (clusters) of interest. Finally, 
phylodynamic analyses are conducted on each cluster of samples independently. 


\subsection{Recent advances}
In the following, we review some computationally efficient approaches for Bayesian phylogenetic inference, including approximate MCMC, online algorithms, and parallel algorithms. While some of the described attempts are promising, they are not yet readily applicable to the type of questions researchers have tried to address in the pandemic. We expect to see many statistical developments in this area in the years to come.

\subsubsection{Sequential Monte Carlo.} 

Sequential Monte Carlo (SMC) methods (also called particle filters) are a set of algorithms used to approximate posterior distributions; See \cite{cho20} for an introduction. SMC-based algorithms have been used to approximate the posterior of phylogenies and mutation parameters through particle MCMC \citep{bouchard2012phylogenetic,wang2015bayesian}. Although in principle these methods can be extended to infer phylodynamic parameters such as EPS, it is not clear how much efficiency can be gained with these methods in comparison to current implementations that rely on Metropolis-Hastings steps. 

Recently, \cite{wang2020annealed} propose to approximate the joint posterior of phylogeny and mutation parameters with a fully SMC approach based on annealed
importance sampling \citep{neal2001annealed}. Here, at each iteration, the SMC algorithm maintains $k$ phylogenetic trees and substitution parameters (particles) with their corresponding weights. The $k$ particles are updated according to traditional Markov chain moves, and acceptance probabilities are based on a likelihood raised to a power (temperature) according to a fixed temperature schedule. 
A great promise of SMC methods is the possibility to be naturally extended to the online setting. We discuss some proposals in the following subsection. 



\subsubsection{Online methods.} \label{subsec:online}
During an outbreak or epidemic, sequencing data often come in sequentially. Redoing the analysis whenever a new sequence becomes available is time-consuming. Thus it is desirable to have an online algorithm that can update the inference using new sequences without having to start the analysis from the beginning. Both \cite{dinh2018online} and \cite{fourment2018effective}, propose online SMC algorithms, which updates the particles and weights when a new sample is added. 
Again, these methods target phylogeny and mutation parameters and requires further work in order to incorporate the SMC approach in phylodynamics.


\cite{gill2020online} propose a distance-based method that adds a new sample to the current sampled phylogeny in the last iteration, simultaneously updating the phylogeny, phylodynamic and evolutionary parameters. The Markov chain is then resumed with the newly added sample. 
This method is applicable to phylodynamic analysis and is implemented in BEAST. \cite{lemey2020accommodating} recently applied this method to update a previous analysis of SARS-CoV-2 sequence data with newly acquired samples.

\subsubsection{Variational Bayes (VB)} 
VB \citep{jordan1999introduction,hoffman2013stochastic,blei2017variational} is a popular alternative to MCMC methods for approximating posterior distributions. Given a class of parametric distributions, VB finds the distribution in the class closest to the target posterior distribution in the sense of Kullback–Leibler (KL) divergence. So the problem of approximating the posterior distribution is recast as an optimization problem, which tends to be faster than classic MCMC. The challenge of applying variational methods to phylogenetics is to choose a sufficiently flexible class of distributions for the tree topologies. \cite{zhang2018variational} introduce a variational algorithm to approximate phylogenetic tree posterior distributions, where the variational family is a class of distributions called the subsplit Bayesian network \citep{zhang2018generalizing}. The subsplit Bayesian network model is defined as the product of conditional probabilities at each internal node (split) from the root to the leaves. The number of parameters of such distribution grows with the number of samples and can potentially be computationally expensive for large sample sizes.
We note that both the SMC and VB methods are only designed for inferring the phylogeny and mutation parameters. It demands further work to apply them to estimate the phylodynamic parameters like effective population size.



\subsubsection{Divide-and-conquer.}

Divide-and-conquer MCMC is an attractive strategy in which the full dataset is partitioned into several subsets; each subposterior---posterior given the subset---is then approximated by running independent MCMC chains, and the subposteriors are then combined to estimate the full posterior \citep{huang2005sampling,neiswanger2013asymptotically,srivastava2015wasp}. However, most of these algorithms rely on the crucial assumption that the subsets are mutually independent. This assumption is violated because molecular sequences share ancestral history (or transmission), modeled by the phylogeny. 
Recently, \cite{liu2021scalable} developed a divide-and-conquer MCMC algorithm, which partitions the dataset into subsets to sample subposteriors separately. They employ a debiasing procedure when combining the posterior samples from the subposterios. 

\section{Testing in phylodynamics}
\label{sec:test}

In the previous section, we described a challenge researchers face while inferring the phylogeny and phylodynamic (coalescent or birth-death) parameters. Inference of these parameters is commonly an intermediate step to address other scientific questions. 

In the current pandemic, we have witnessed a surge of novel variants that have caused public health concern \citep{vol21nat,dav21}. A significant focus of SARS-CoV-2 research has been the study of whether specific mutations (variants) impact viral properties, such as transmissibility, virulence, and the ability to increase disease severity. While it is often possible to study cell infectivity in animal models and to study \textit{in vitro} whether a mutation is associated with changes in viral phenotypes, determining whether it leads to significant differences in viral transmission or disease response relies on observational data from both, the pathogens and the hosts. These data often consist of molecular sequences, epidemiological and clinical data. These types of statistical analyses are challenging because although an increase in frequency is a signal of selective advantage, observed increase can also be the product of many other factors such as multiple introductions and human behaviors. In this section, we restrict our attention to two types of analyses designed to test whether there are significant differences in transmissibility between a variant of concern (VOC) and a non-VOC. The first type is solely based on molecular data, and the second type utilizes molecular data paired with phenotypic traits and clinical data.


\subsection{Detecting higher transmissibility relying solely on molecular data}
\label{sec:phylo}

\subsubsection{Practices in analyzing SARS-CoV-2 data.} 

Phylodynamic models have been applied to estimate the effective population size (EPS) of several VOCs and non-VOCs from molecular samples. It is assumed that the non-VOC spread through the population and accumulated variation before the VOC appeared in the population. If the VOC confers higher transmissibility, the EPS of VOC should increase at a higher rate than the non-VOC in multiple locations around the world. Moreover, comparisons between the two EPSs should be based on VOC and non-VOC samples sharing the same environmental factors such as public policies and temporal seasons to control for possible confounders. 

\cite{vol21} stress the need to observe repeated independent introductions of each variant and follow their trajectories. The authors analyzed molecular data collected in the UK during the first six months of 2020 to test whether the VOC (D614G substitution) had selective advantages. The authors first obtained the MLE phylogeny of samples available in the UK, and used it to identify several clusters (VOC clusters and non-VOC clusters). These clusters included one or a small number of introductions of the virus in the UK. The authors then estimated EPS growth rates for each cluster and compared the posterior distributions of VOC and non-VOC growth rates.
They observed that the two distributions are largely overlapped and thus there is no significant difference.

Other studies have also identified multiple introductions for estimating VOC growth rates. For example, \cite{dav21} considered introductions across different countries. A challenge with this type of analyses lies in the detection of independent introductions. It is unclear how ignoring phylogenetic uncertainty affects the definition of introductions and estimation of EPS, and whether introductions in different locations can be treated as independent.   
\cite{vol21nat} performed a phylodynamic case-control study which consisted in selecting 100 random samples of 1000 sequences with the VOC (alpha variant) paired with another 1000 non-VOC sequences. Those sequences were matched by the week and the location of the collection. The random samples were selected with weights proportional to the number of reported cases per week and local authority in the UK and hence, expected to be representative of the UK. The 200 phylogenies were estimated via MLE, and 200 EPS trajectories were inferred from each tree in order to obtain two bootstrap distributions of VOC and non-VOC EPSs. In their study, the comparison of the two EPS distributions supported an increase in the transmissibility of the VOC. 

A simple and popular strategy to estimate the growth rate of the VOC and non-VOC populations consists of modeling the sampling times of sequences solely (ignoring molecular data) through a logistic growth model \citep{vol21,vol21nat,dav21,tru21}. Data, in this case, consists of counts of genomes belonging to the VOC and the non-VOC over time, with counts binned into weeks. This type of analysis simply models the proportion of VOC sequences over the total number of collected sequences over time, rather than looking at the EPS.  

In the two phylodynamic studies discussed, the EPSs are estimated independently for the two populations. We argue that this approach might be suboptimal, not only because the two trajectories may be correlated but also because the uncertainty quantification in the difference between growth rates can be somewhat lost in the aggregation step. In the next section, we discuss a simple hierarchical model that jointly models the two lineages so that the difference in growth rates is easily interpretable.



\subsubsection{A simple model to test for population growth}
\label{sec:new_mod}

Assume that we are provided with the two phylogenies $\bfg_{0}$ and $\bfg_{1}$ of the non-VOC and the VOC, respectively. We can model the two phylogenies as conditionally independent given a shared baseline EPS denoted by $N_{e}(t)$. More specifically, we assume $\bfg_{0}$ is a realization of a CP with parameter $N_{e}(t)$ and $\bfg_{1}$ is a realization of a CP with parameter $\alpha N_{e}(t)^{\beta}$. Here, $\alpha>0$ is a scaling parameter, and $\beta$ is the main parameter of interest: $\beta=1$ indicates that the growth rate of the EPS in the two groups is identical, $\beta>1$ indicates that the growth in genetic diversity of the VOC is larger than the non-VOC.
Note that $\beta$ can also take negative values.

This simple model is highly interpretable, with a single parameter, $\beta$, quantifying the change in transmissibility of the VOC relative to the non-VOC. 
One can choose the preferred prior on $N_e(t)$, such as a Gaussian Markov random field (GMRF) \citep{min08}, a Gaussian process \citep{pal13}, and the Horseshoe Markov random field \citep{fau20}. The posterior distribution $P(N_e(t), \alpha,\beta \mid \bfg_{0},\bfg_{1})$ can be computed in a few seconds with INLA \citep{rue09}. The accuracy of this approximation in phylodynamics has been studied in \cite{lan2015efficient}. We provide a publicly available implementation of the following model in \texttt{phylodyn} available in {\tt https://github.com/JuliaPalacios/phylodyn}:

\begin{equation}
	\label{eq:sel}
	\begin{array}{lll}
		\bfg_{0} \mid N_e, \mathbf{y}_{0} &\sim& \text{Coalescent with EPS} \quad N_e(t) \\
		\bfg_{1} \mid N_e, \alpha, \beta,\mathbf{y}_{1} &\sim& \text{Coalescent with EPS} \quad \alpha N_e(t)^{\beta} \\
		\log N_e &\sim& \text{GMRF of order 1 with precision} 
		\quad \tau\sim \text{Gamma}\\
		\log\alpha \mid \sigma_0^2 &\sim&  \mathcal{N}(0, \sigma_0^2), \\
		\beta \mid \sigma_1^2 &\sim&  \mathcal{N}(0, \sigma_1^2),  \\
	\end{array}
\end{equation}

Model \eqref{eq:sel} enforces a strict parametric relationship between the two EPSs. 
While this may be too restrictive, we argue that it is a reasonable price to pay for the sake of interpretability and parsimony. We illustrate the methodology by applying the model to SARS-CoV-2 sequences collected in Washington state at the beginning of the epidemic.\\

\begin{figure}
	\centering
	\includegraphics[width=0.5\textwidth]{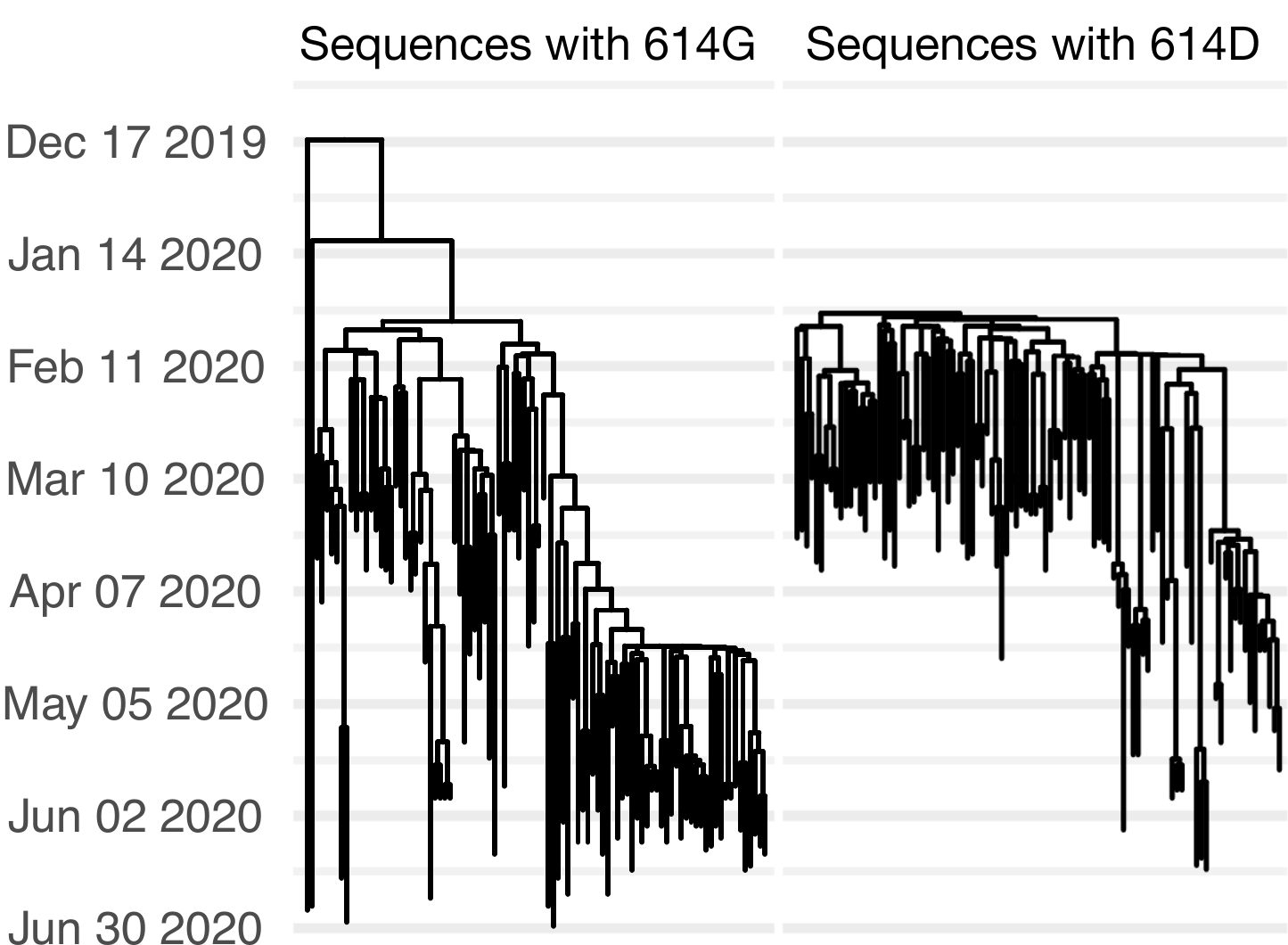}
	\caption{\small{\textbf{Phylogenies of the G and D variants inferred in Washington state.} Phylogenies are the maximum clade credibility trees obtained from posterior distributions estimated with BEAST (Appendix). Each tree is generated from $100$ sequences chosen at random among those collected in Washington state between January 1, 2020 to June 30, 2020. The left tree includes sequences with G in the codon position 614 of the viral spike protein. The right tree includes sequences with D in the codon position 614. By 2021, the G type dominated the pandemic.}}
	\label{fig:tree}
\end{figure}

\begin{figure}
	\centering
	\includegraphics[width=1\textwidth]{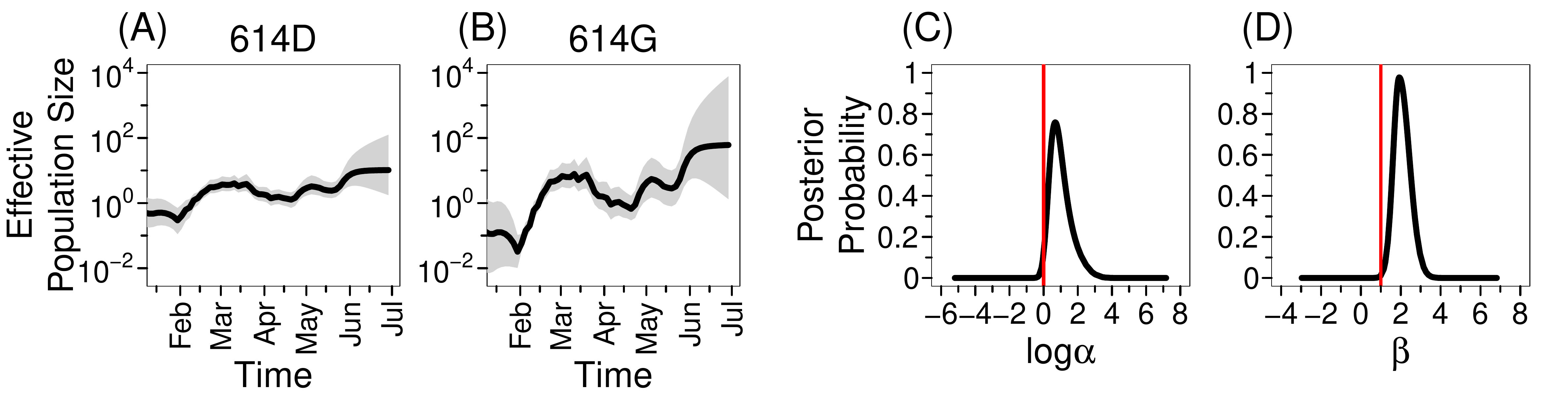}
	\caption{\small{\textbf{Effective population sizes of D and G variants in Washington State.} Panels (A-B) depicts posterior mean of $Ne(t)$ and $\alpha N_{e}(t)^{\beta}$, the effective population size trajectories of the $D$ and the $G$ variants respectively. Shaded areas represent 95\% BCIs. Panel (C) depicts estimated posterior distribution of $\log \alpha$ and panel (D) depicts estimated posterior distribution of $\beta$. Red lines indicate the values of $\log \alpha$ and $\beta$ under the hypothesis that both variants share the same effective population size trajectory.}}
	\label{fig:posterior}
\end{figure}

\noindent \textit{Application to SARS-CoV-2 sequences in Washington state.} We randomly selected 100 sequences with the D codon (non-VOC) and 100 sequences with the G codon in position 614 (VOC), from the 4356 publicly available sequences in GISAID \citep{shu2017gisaid} collected in Washington state between January 1, 2020 and June 30, 2020. We analyzed the two samples independently and obtained the two phylogenies (Figure \ref{fig:tree}) by summarizing the two corresponding posterior distributions obtained with \texttt{BEAST2} \citep{bou19}. Details of model and MCMC parameters are located in the Appendix.



Panels (A-B) of Figure~\ref{fig:posterior} depict the posterior medians (solid lines) and 95\% BCIs of $N_{e}(t)$ (shaded areas) obtained by fitting model \ref{eq:sel}. Panel (C) depicts the estimated posterior distribution of  $\log \alpha$ with posterior mean close to 0 and panel (D) depicts the estimated posterior distribution of $\beta$. In the random subsample considered, sequences with the D variant generally have an earlier collection date than sequences with the G variant. This is consistent with the general observation that the G variant progressively replaced the D variant \citep{nextstrain}. The main parameter of interest is $\beta$. The posterior distribution is unimodal, with mean $2.08$ and $95\%$ credible region $(1.35,2.97)$. It is well above $1$, suggesting that EPS growth is more pronounced among sequences having the G variant. The impact of $\beta \approx 2$ is evident in the first two panels of Figure \ref{fig:posterior}, where the EPS of G grows at a higher rate than that of the control group.




A benefit of the model described here is the flexible nonparametric prior placed on $N_e$. Panels (A-B) of Figure~\ref{fig:posterior} suggest that parametric models would not reasonably approximate the trajectory: for this dataset, our estimates indicate that $N_e$ fluctuates in the period considered. The goal of the analysis is inferring the parameter $\beta$. Hence, we argue that the best possible fit in modeling $N_e$ is necessary. A future development includes the inference of the proposed model parameters from molecular data directly.

\subsection{Combining molecular sequence data and other types of data}
\label{sec:treetest}



We now examine the situation when viral molecular data are matched with host clinical data and we are interested in testing an association between clinical traits such as disease severity and transmission history. For example, does the variant of concern affect disease severity? We first describe some  current practices in testing for such associations.

\subsubsection{Practices in analyzing SARS-CoV-2 data}

If one is studying a VOC, the variant naturally partitions the hosts into two groups: individuals carrying the VOC and those carrying the non-VOC. Here, one can resort to standard statistical tests for detecting changes in mean or distribution \citep{vol21,vol21nat,dav21,leu21}. For example, \cite{vol21} study the mutation in codon position 614 (D and G mutations) and analyze the difference in several response variables such as disease severity and age using a Mann-Whitney U-test. One related approach that tests for overall correlation between phenotypes and shared ancestry (transmission structure) that accounts for phylogenetic uncertainty is the BaTS test \citep{parker2008correlating}. This method relies on simple statistics such as the parsimony score and association index. 

However, the situation is more challenging when the candidate VOC has not yet been identified. For example, \cite{zha20} estimated a phylogeny and used it to identify two major clades, the authors then characterized these two clades \textit{e.g.} which mutations differentiate them, and tested for association with clinical data. Here the choice of which clades to pick and compare is somewhat arbitrary. 

\subsubsection{Recent advances.}
\cite{beh20} recently proposed treeSeg, a method for testing multiple hypotheses of association between a response variable and the phylogeny tree structure. A key feature in treeSeg is to formulate the testing problem as a multiscale change-point problem along the hierarchy defined by a given phylogeny. The test statistic is based on a sequence of likelihood ratio values, and the change-point detection methodology is based on the SMUCE estimator \citep{fri14}. This method was recently applied to test an association between the inferred phylogeny from SARS-CoV-2 sequences collected in Santa Clara County, California in 2020, and disease severity \citep{par21}. The authors did not find any significant association. 

One statistical challenge in applying treeSeg to phylodynamics is that it ignores uncertainty in the tree estimation. If a subtree is found to have an association to the response, we can assess uncertainty in the subtree formation (independent of treeSeg analysis) by an estimate of the subtree posterior probability or the subtree bootstrap support \citep{efron1996bootstrap}. A more integral approach is an open problem. 

Another situation arises when we are interested in assessing phenotypic correlations among traits \citep{fel85,gra89,pag94}. 
Here, multiple traits are modeled as stochastic processes evolving along the branches of the phylogeny; for example, as Markov chains \citep{pag94}, or as multivariate Brownian motion \citep{fel85,hue03,fel05,fel12,cyb15}. 
Despite their relevance in understanding viral evolution and drug development, computation is the main limitation preventing the widespread use of this methods' class. \cite{zha21} is a recent attempt to make inference more scalable. They introduce an algorithm based on recent advances in the MCMC literature (the Bouncy particle sampler \citep{bou18}). Although, the implementation of the methodology seems quite involved, somewhat preventing broader applicability. 

\section{Preferential sampling}
\label{sec:pref}

In the standard CP, the temporal sampling process of sequences is assumed to be fixed and uninformative of model parameters. However, the sampling process that determines when sequences are collected can depend on model parameters such as the EPS in some situations. In spatial statistics, preferential sampling arises when the process that determines the data locations and the process under study are stochastically dependent \citep{dig10}. In coalescent-based inference, this can be incorporated by modeling the sampling process as an inhomogeneous Poisson process (iPP) with a rate $\lambda(t)$ that depends on $N_e(t)$. If the model is correct, the sampling times can provide additional information about the EPS $N_{e}(t)$. The statistical challenge is that when the model is misspecified, incorrectly accounting for preferential sampling can bias the estimation of the EPS. The same situation occurs in the BDSP in which the sampling process depends on the death rate \citep{Stadler2010,Volz2014,cap21}.

\subsubsection{Recent Advances.}
Table \ref{tab:prefmethdo} lists different models and implementations that account for preferential sampling in phylodynamics. Among the parametric approaches, \cite{Volz2014} model the EPS $N_e(t)$ as an exponentially growing function and $\lambda(t)$ is linearly dependent on the EPS.  \cite{kar16} assume that $N_e(t)$ is a continuous function modeled nonparametrically with Gaussian process priors, and $\lambda(t)=\exp(\beta_0)N_e(t)^{\beta_1}$, for $\beta_0,\beta_1\geq0$, i.e. the dependence between the sampling process and the effective sample size is described by a parametric model. While this model is computationally appealing, the strict parametric relationship between the sampling and coalescent rates can induce a bias if the sampling model is misspecified (see simulations in \cite{cap21}).

\cite{par20} propose an estimator called Epoch skyline plot, that allows the dependence between the rate of the sampling process and $N_{e}(t)$ to vary over time. In \cite{par20}, $\lambda(t)$ is a linear function of $N_e(t)$ within a given time interval, but the linear coefficient changes across time intervals. This framework allows practitioners to incorporate heterogeneity in the sampling design over time. \cite{cap21} extends this approach, modeling both $N_e(t)$ and $\lambda(t)$ nonparametrically, employing Markov random field priors on both $N_e(t)$ and a time-varying coefficient $\beta(t)$. Here, the dependence between the two processes is modeled through $\lambda(t)=\beta(t) N_e(t)$. \cite{cap21} show through simulations that the more flexible dependence between the sampling and the coalescent processes the less the risk of biasing the $N_e(t)$ estimate because of model misspecification while still retaining the advantages of the parametric approaches (narrower credible regions). \cite{kar20b} assume that $\lambda(t)=\exp(\beta_0)N_e(t)^{\beta_1} +\boldsymbol{\beta}'\textbf{X}(t)$, where $\textbf{X}$ is a vector of covariates and $\boldsymbol{\beta}'$ the corresponding linear coefficients. Here, a covariate can be a dummy variable indicating the implementation of lockdown measures. The covariate-dependent preferential sampling requires the availability of information related to the sampling design. Finally, the methodologies of \cite{kar16} and \cite{cap21} rely on a know phylogeny, while \cite{Stadler2010},\cite{par20}, and \cite{kar20b} account for uncertainty in the phylogeny.\\

\begin{table*}
	\caption{Approaches for preferential sampling}
	\label{tab:prefmethdo}
	\begin{tabular}{@{}lll@{}}
		\hline
	Method & Implementation & Author \\
		\hline
	Birth-death & \texttt{BDSKY} (BEAST2) & \cite{Stadler2010}\\
	Parametric $N_e$ and $\lambda(t)$ & $NA$ & \cite{Volz2014} \\
	Nonparametric $N_e$ and $\lambda(t)=\exp(\beta_0)N_e(t)^{\beta_1}$ & \texttt{phylodyn} (R package) & \cite{kar16}\\
Epoch Skyline plot (Nonparametric $N_e$ and $\lambda(t))$)	& \texttt{BESP} (BEAST2) & \cite{par20}\\
adaPref (Nonparametric $N_e$ and $\lambda(t)=\beta(t)N_e(t)$) & \texttt{adaPref} (R package) & \cite{cap21}\\
	Nonparametric $N_e$ and $\lambda(t)=\exp(\beta_0)N_e(t)^{\beta_1} +\boldsymbol{\beta}'\textbf{X}(t)$ & BEAST & \cite{kar20b}\\
		\hline
	\end{tabular}
\end{table*}

\noindent \textit{Application to SARS-CoV-2 sequences in Washington state.} We continue the analysis of the Washington molecular sequences introduced in Section~\ref{sec:new_mod}. We infer the EPS of the two groups (sequences with 614G and sequences with 614D) from the phylogenies inferred with BEAST2 and plotted in Figure \ref{fig:het_tree}. We compare three different models: one that ignores preferential sampling \citep{pal12}, the parametric preferential sampling model of \cite{kar16}, and the adaptive preferential sampling of \cite{cap21}. All three models share a GMRF prior on $N_e(t)$.

\begin{figure}
	\centering
	\includegraphics[scale=0.4]{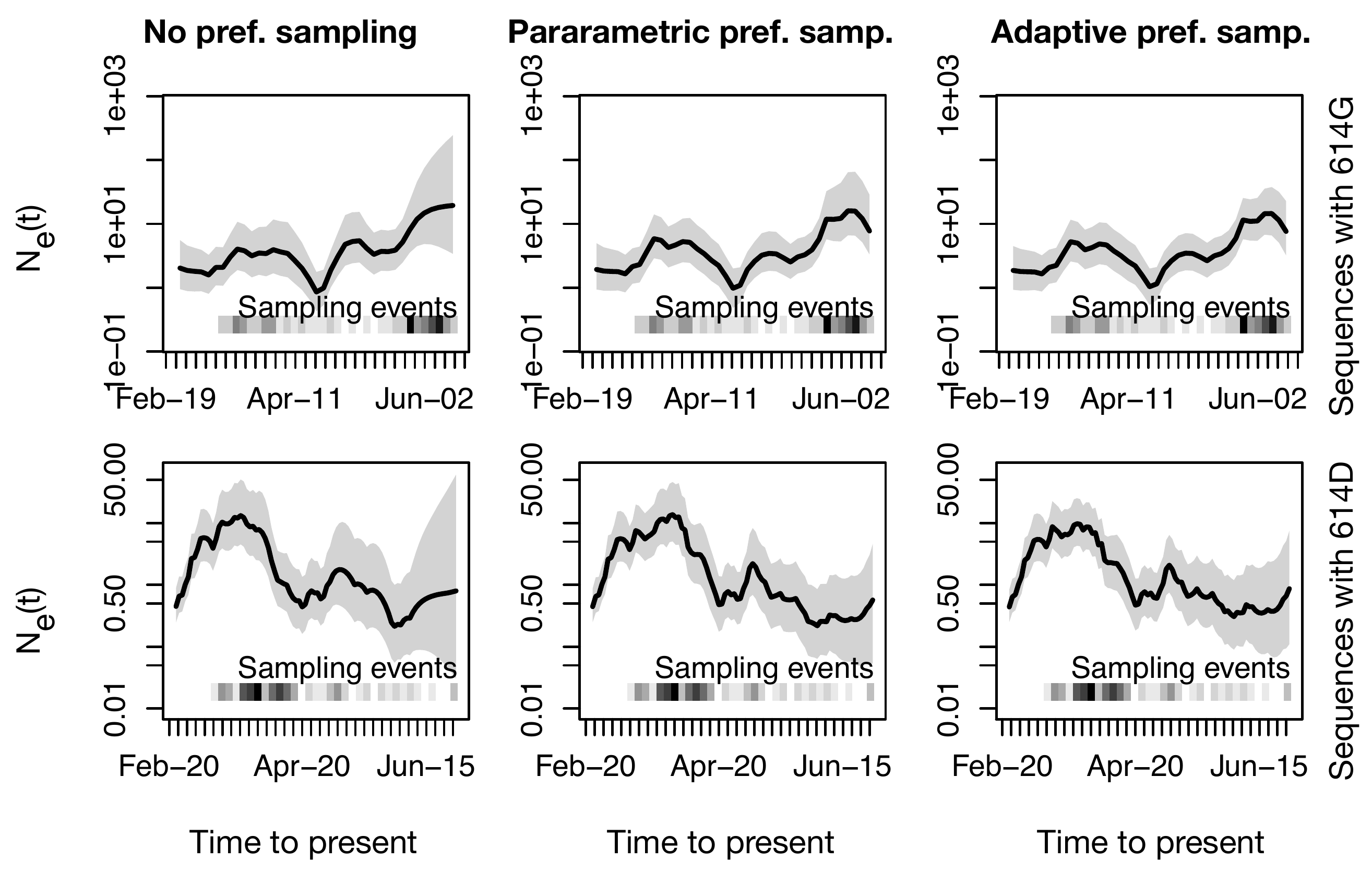}
	\caption{\small{\textbf{$N_e(t)$ estimated from SARS-CoV-2 phylogenies of sequences from Washington state.} The first row depicts EPS of the G type in codon position 614, the second row depicts EPS of the D type in codon position 614. The first column estimates are obtained with model of \cite{pal12} that ignores preferential sampling, the second column with the model of \cite{kar16} that parametrically models preferential sampling, and the third column models with the adaptive preferential sampling model of \cite{cap21}. In each panel, black lines depict posterior medians and the gray areas the $95\%$ credible regions of $N_{e}(t)$. Sampling times are depicted by the heat maps at the bottom of each panel: the squares along the time axis depict the sampling time, while the intensity of the black color depicts the number of samples.}}
	\label{fig:pref_samp}
\end{figure}

Figure~\ref{fig:pref_samp} depict the EPS posterior distributions obtained with the three methods applied to the two genealogies. At the bottom of each panel, heat maps represent the sampling times (intensity of the black color is proportional to the number of samples collected.) The estimates of the two models accounting for preferential sampling are pretty similar, while not modeling the sampling process leads to a slightly different population size trajectory.

As expected, including a sampling process reduces the credible region width: the mean width of the $95\%$ credible region is much wider for the model that ignores preferential sampling in the sequences with 614G with respect to any of the models accounting for preferential sampling (approximately $6$ times large in the first two, and $3.5$ times in the second row).

Under the preferential sampling assumption, the more sequences are collected, the higher the EPS is. The effect is evident in Figure~\ref{fig:pref_samp}. For example, in the month of June, we see that the EPS grows both for the sequences with 614G and 614D if we ignore sampling information. If we account for preferential sampling, the EPS of sequences with 614G ``dips" because no sequences in the last week were part of our dataset. 

This application offers a cautionary tale on this class of models.  Modeling the sampling process not only reduces the credible region width, it can also affect the estimates heavily. This behavior signals that ignoring the sampling process leads to a bias if the preferential model is correctly specified. The opposite is also true: we could be biasing our estimates if the sampling model is incorrect by including sampling times information.


\section{Phylodynamic inference of epidemiological parameters} \label{sec:sir}
Epidemiological parameters are often estimated from case count time series; these estimates, however, can be biased due to delays and errors in reporting. Sequence data provide complementary information that can be used for estimating critical epidemiological parameters within a phylodynamic framework. Formal model integration of the CP and epidemiological compartmental models establishes a link between the EPS of pathogens and the underlying number of infected individuals. Equivalently, in the forward-in-time BDSP model, parameters such as the rate of transmission and effective reproduction number can be directly inferred from molecular data. 


\subsection{Phylodynamic inference relying solely on molecular data}

\subsubsection{Phylodynamic inference with CP-EPI.} 
While a linear relationship between the viral EPS and the disease prevalence exists at \textbf{endemic equilibrium}, such simple correspondence is not valid in general \citep{Koelle2012}. The CP-EPI provides a probability model of a phylogeny in terms of epidemiological parameters by linking the EPS trajectories to a mechanistic epidemic model \citep{Volz2009}. The infectious disease population dynamics can be modeled as a CTMC whose state space is the vector of occupancies in compartments corresponding to disease states. However, the transition probability becomes intractable even for the simplest SIR model \citep{Tang2019}. One way to mitigate the computational issue is to deterministically model the disease dynamics \citep{Kermack1927}; we term such model as a deterministic CP-EPI.

\citet{Volz2009} developed a theoretical basis for the deterministic CP-EPI. In the particular case of SIR dynamics, the population is divided into compartments. At time $t$, the state is $\{S(t), I(t), R(t)\}$, of susceptible, infected and recovered individuals respectively. In this context, the phylogeny represents the ancestry of a sample of infected individuals in the population. Let $A(t)$ denote the number of lineages ancestral to the sample in the phylogeny at time $t$. The probability that a transmission event at time $t$ corresponds to a transmission event ancestral to the sample is $\binom{A(t) } {2}/ \binom{I(t)} {2}$. Denoting the total number of new infections at time $t$ by $f(t)$, the rate of coalescence is


%
\begin{equation} \label{eq:volz.orig}
	\lambda_A(t) = f(t)\frac{ \binom{A(t)} {2 }}{\binom{I(t)} { 2}} \approx \binom{A(t)} {2}\frac{2f(t)}{I^2(t)}, \hspace{5pt} 
\end{equation}
Assuming a per capita transmission rate $\beta$, $f(t) = \beta S(t)I(t)$ is  the number of transmissions per unit time (the incidence of infection). The population dynamics of compartments, $\{S(t), I(t), R(t)\}$, is governed by an initial state and a system of ordinary differential equations. Recall that $\lambda_A(t) = \binom{A(t)} { 2} / N_e(t)$ is the coalescence rate in the standard CP, we then get
\begin{equation} \label{eq:volz.orig.2}
N_e(t) = \frac{I^2(t)}{2f(t)}.
\end{equation}

The initial CP-EPI model has been extended to incorporate serial sampling, population structure, time- and state-dependent rate parameters, and a large class of epidemic processes  \citep{Volz2012, Volz2018}. In \citet{Volz2018}, the authors assumed that recovery rate and number of susceptible individuals are known; the transmission rate is modeled as a straight line with normal prior on the slope parameter and lognormal prior on the intercept parameter. Inference is performed via MCMC in BEAST2 \citep{bou19}. We note that in the SIR models, not all parameters are identifiable. We usually need to assume known values of some parameters and very informative priors (See \cite{louca2021fundamental} for further details).

The deterministic CP-EPI has provided a computationally efficient framework for studying the evolution and pathogenesis of SARS-CoV-2 via estimating $R_0(t)$ at the beginning of the pandemic \citep{Volz2020, Geidelberg2021}, fine-scale spatiotemporal community-level transmission rate variation \citep{Moreno2020}, and the effects of control measures on epidemic spread \citep{Miller2020, RagonnetCronin2021}. 


So far, we have ignored within-host evolution, that is, we have assumed that pathogen diversity within a host is negligible. It can be shown that Equation~\ref{eq:volz.orig} is a limiting case of a more general model \citep{Dearlove2013, Volz2017}, which relaxes many assumptions from the previous derivation, such as negligible evolution within host. In the metapopulation CP-EPI, which is based on the metapopulation CP \citep{Wakeley2001}, each deme corresponds to a single infected host and can be reinfected more than once. Within each host, there is a non-negligible pathogen population size, and the within-host coalescence does not occur immediately following an infection. Further, during an inter-host transmission, non-negligible genetic diversity can be transmitted across hosts. Due to its complexity, the current metapopulation CP-EPI model assumes constant rate parameters and deterministic disease dynamics. 

As empirical evidence of reinfection and of the effects of within-host diversity on patient disease severity and transmissibility mounts for SARS-CoV-2 \citep{Tillett2021,AlKhatib2020, San2021}, it is becoming apparent that developing computationally tractable methods that incorporate both time-varying parameters and stochasticity into the metapopulation CP-EPI framework is an important future direction in the field.

While the deterministic CP-EPI is computationally efficient, epidemiological dynamics are inherently stochastic, with both demographic and environmental stochasticity playing important roles in disease dynamics. The deterministic epidemic model can lead to overconfident estimations when the disease prevalence is low or the population size is small, or when fitting models to long-term data, as the effects of stochasticity accumulate over time \citep{Popinga2014}.
The CP-EPI with the stochastic epidemic model, which we term as the stochastic CP-EPI, is better suited for addressing important epidemiological questions, such as the early-stage behavior of an epidemic, the outbreak size distribution, and the extinction probability and expected duration of the epidemic, while accounting for the uncertainties in the estimations \citep{Britton2010}. 

\subsubsection{Phylodynamic inference with BDSP.}

The BDSP \citep{Stadler2013a} introduced in Section~\ref{subsed:intro.bd} has been extended to incorporate population structure \citep{Kuehnert2016} and it has been applied for inferring $R_e(t)$ early in the SARS-CoV-2 pandemic in Europe \citep{Nadeau2021, Hodcroft2021}. 
The BDSP, however, requires specification of the sampling probabilities,
and its misspecification results in biased estimates, as demonstrated in inferring $R_0$ from the SARS-CoV-2 data in the northwest USA \citep{Featherstone2021}. This is because, in the BDSP, sampling times provide information about the whole population dynamics \citep{Volz2014}.

An important distinction between BDSP models and the CP-EPI models is that the BDSP model is parameterized in terms of birth, death, and sampling rates, however it does not directly model the number of infected individuals over time and the number of recovered individuals over time. The CP-EPI, instead, directly models the number of individuals in each compartment, together with birth and death rates. We note that the BDSP has been extended to infer the prevalence trajectory from molecular sequences and case count data (Section~\ref{sec:phy_count_2}).

\subsection{Phylodynamic inference relying on molecular data and disease count data}

When fitting mechanistic population dynamic models, integrating multiple sources of information, particularly time series surveillance data, with molecular sequence data, can improve phylodynamic inference of epidemic model parameters. This subsection describes extensions of CP-EPI and BDSP to incorporate both data sources.

\subsubsection{Phylodynamic inference with CP-EPI}
 \citet{Rasmussen2011} employed PMCMC for Bayesian inference under the stochastic CP-EPI, from both a fixed phylogeny and time series incidence data. In their implementation, they allow the transmission and recovery rates to vary in time. Unfortunately, inference is computationally expensive due to the high-dimensional parameter space. Other extensions to this framework include incorporation of overdispersion in secondary infections \citep{Li2017}. Recently, \citet{Tang2019} proposed to bypass PMCMC and used a linear noise approximation. The authors approximated the SIR transition density with a Gaussian density and developed an MCMC algorithm for this approximate inference. 


Current implementations of the stochastic CP-EPI have a few limitations, many of which stem from computational cost. This reduces their utility in SARS-CoV-2 analyses. First, most methods have adopted an epidemic model with one infection compartment and ignore further population structure, such as spatial distribution and age. Second, statistical dependency between sampling times and latent prevalence is ignored. If the sampling process is known, we could incorporate sampling model directly as in the preferential sampling \citep{kar16}, for improving parameter estimation ; see Section~\ref{sec:pref}. Finally, to fully account for phylogenetic uncertainty, a computationally efficient method for directly fitting stochastic epidemic models to genetic sequences will be needed.



\subsubsection{Phylodynamic inference with Birth-Death processes.}\label{sec:phy_count_2}
There have been a few recent developments in joint modeling of molecular data and case count records under the birth-death population dynamics. Recently, \citet{Gupta2020} extended the BDSP model \citep{Stadler2010}  to include case count data. The authors derive the density of the phylogeny jointly with case count data in terms of the BDSP rates. This work was later extended to model prevalence \citep{Manceau2021}; building on \citet{Gupta2020}, the authors derived the density of the prevalence trajectory conditioned on the phylogeny and case count data. Finally, \citet{Andreoletti2020} extended this work to allow for piecewise constant rates and used it to estimate $R_e$ and prevalence of the SARS-CoV-2 Diamond Princess epidemic that occurred in Jan-Feb 2020. 

\cite{Vaughan2019} recently proposed a method that differs from the BDSP discussed in the previous paragraph. The authors propose to estimate the posterior distribution of the full epidemic trajectory, together with the phylogeny and model parameters from molecular sequence data and count data. Here, the authors express the density of the phylogeny jointly with case counts, conditionally on the full epidemic trajectory that consists of the sequence of events (infection, sampling and recovery) and their corresponding event times. The posterior distribution is estimated with PMCMC.

\subsection{Phylodynamic inference with approximate Bayesian computation and deep learning}
As SARS-CoV-2 continues to spread, the virus is subject to strong host and \textbf{anthropogenic selective} pressures as has already been exemplified by the emergence of the new variants exhibiting adaptive \textbf{antigenic evolution} \citep{Zhou2021, LopezBernal2021}. 
As discussed in Section~\ref{sec:scalability}, however, the likelihood-based Bayesian computation methods are computationally expensive, prohibiting the application of more complex and realistic phylodynamic models such as those involving structured populations, natural selection and recombination. 
To overcome this obstacle, likelihood-free rejection sampling methods based on approximate Bayesian computation (ABC) \citep{Beaumont2002} have been developed for phylodynamic studies. The phylodynamic ABC methods \citep{Ratmann2012, Poon2015, Saulnier2017} first simulate a large number of phylogenies under complex epidemiological models with different parameter values, then quantify the discrepancy between simulated and ``observed'' phylogenies and accept the ones close to the target to construct an approximate posterior distribution of the model parameters. Here, the ``true'' phylogeny is unknown and an estimated phylogeny from sequence data is used as the ``observed'' phylogeny. The phylogenetic dissimilarity measure can be either a function of summary statistics \citep{Saulnier2017}, where each extracts a specific feature of the phylogeny, or a metric defined directly on the space of phylogenies \citep{Robinson1981, Billera2001, Colijn2017, Kim2020}. To improve computational efficiency, \citet{Ratmann2012} and \citet{Poon2015} used ABC-MCMC \citep{Marjoram2003}, while \citet{Saulnier2017} employed regression-based ABC \citep{Blum2010}. 

The ABC-based methods, however, are known to be sensitive to the choice of summary statistics, similarity measures, and match tolerance \citep{Lintusaari2016}. As an alternative, \citet{Voznica2021} proposed a rejection-free approach for estimating epidemiological parameters and for model selection based on deep learning: feed-forward neural network (FFNN) with a large set of summary statistics that were curated for phylodynamic regression-ABC \citep{Saulnier2017} and convolutional neural network (CNN). A key component in their method is their proposed bijective encoding of (unlabeled) phylogenetic trees as vectors, amenable to standard deep learning methods. As the framework assumes a known phylogeny as an input, phylogenetic uncertainties are not accounted for. While the computational burden lies in simulating the training data and training the network, once trained, the parameter estimation is very efficient without having to retrain the model with new data. They show comparable accuracy under the basic BDSP model and better accuracy under more complex models, which incorporate factors such as superspreader events, when compared to the current popular likelihood-based methods. As the number of SARS-CoV-2 sequences grow exponentially and its disease dynamics varies across regions, the deep learning framework can offer a fast alternative for monitoring the epidemic.

\section{Discussion}
\label{sec:disc}

Statistical methods in molecular epidemiology offer powerful tools to help us understand and monitor a pandemic as it unfolds. Our paper has outlined some of the statistical models used for tracking SARS-CoV-2 and identified a few areas where state-of-the-art phylodynamic approaches fell short of delivering their full potential. 

The lack of scalable inference methods that can analyze the unprecedented amount of molecular sequences available is a common theme among all SARS-CoV-2 analyses discussed here. Popular strategies include subsampling, inferring a fixed phylogeny and using a fixed phylogeny for partitioning the data. It is generally missing how stable the results are to these choices. We chose to evaluate the stability of phylogenetic posterior from subsampling. 

\begin{figure}
	\centering
	\centering
	\includegraphics[width=0.7\textwidth]{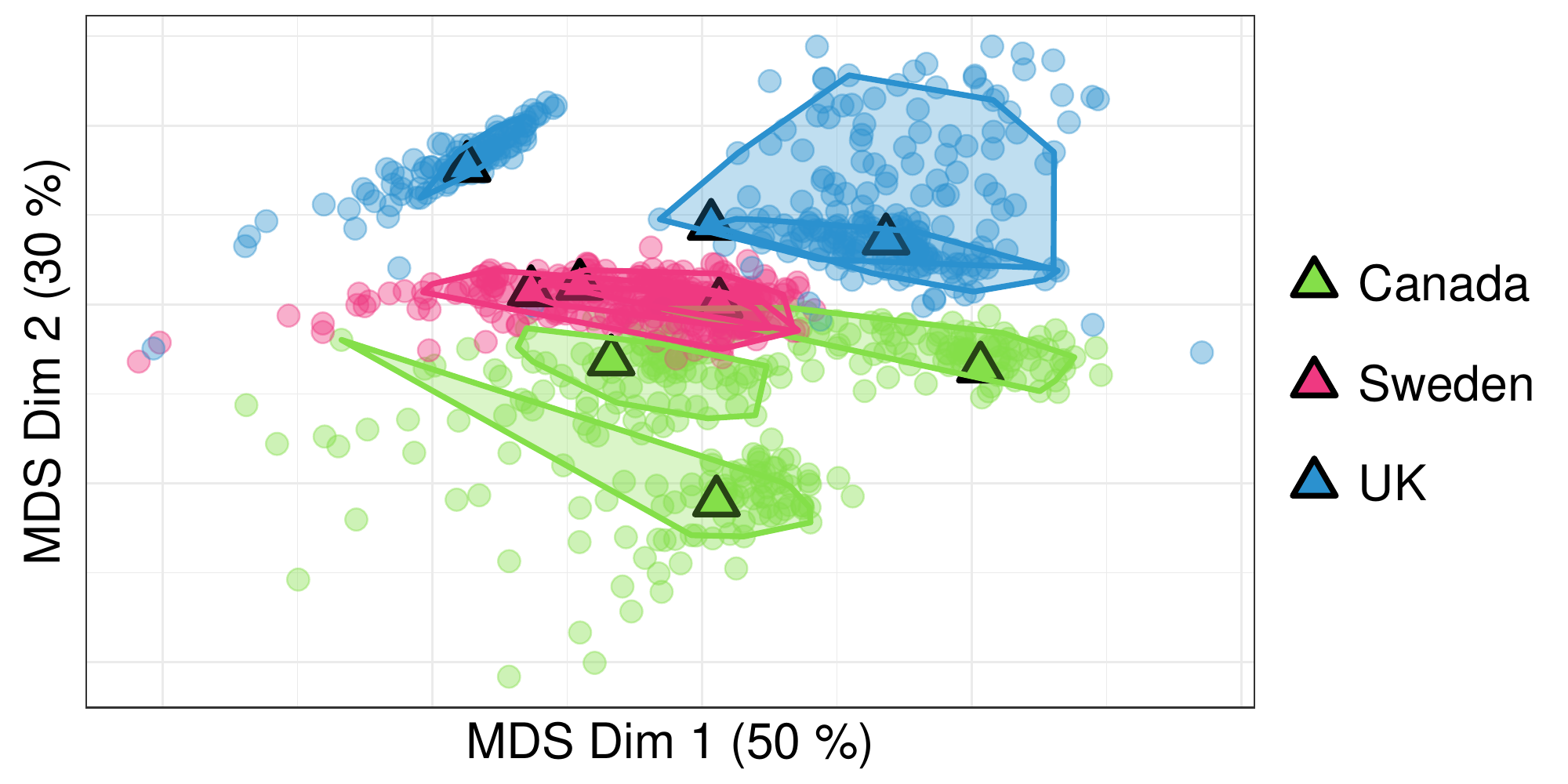}
	\caption{\small{\textbf{MDS plot of phylogenetic posterior distributions from Canada, Sweden, and UK.} Each dot represents a phylogeny from one of the nine posterior distributions (three distributions per country). Each posterior (cluster of phylogenies) is estimated from 100 samples randomly chosen from November 2020 to February 2021. Metric on tree spaces given by \cite{Kim2020}. The triangles indicate the medoids of each distribution and the shaded regions corresponds to 50\% credible convex hulls around the medoids.}}
	\label{fig:mds}
\end{figure}

\cite{raj21} proposed a visual inspection of several phylogenetic posteriors obtained from different samples, to investigate phylogenetic stability. If the distributions overlap, then there is indication of phylogenetic stability. We followed the proposed methodology to investigate the stability of the phylogenetic posteriors of SARS-CoV-2 obtained from Canada, Sweden and the UK. We took three samples, each containing 100 sequences chosen at random from each country available in GISAID, from November 1, 2020 to February 1, 2021. The nine posterior distributions are projected in two dimensions and depicted in the MDS plot of Figure \ref{fig:mds}. Here, Sweden is the only country that shows phylogenetic stability. We recommend performing stability analyses when choosing a small subset of samples.

We centered our scalability discussion on Bayesian algorithms that either aim to replace MCMC by sequential Monte Carlo or variational inference, or to potentially improve the convergence of MCMC. Other approaches include the online strategy, which updates of the posterior distribution as sequences become available sequentially, as well as the divide-and-conquer strategy, which divides the data into smaller subsets. We anticipate several advancements in this area in the future. 

Apart from Bayesian computation, an important direction for improving scalability includes more efficient phylogenetic modeling. As the number of samples increases, the probability of observing sequences with identical genotypes also increases. 
Phylogenies with permuted labels of samples with identical genotypes have equal likelihood.
States of the CP can then be lumped together in a situation like this. These lower resolutions of the coalescent have smaller cardinality and can potentially be more efficient \citep{sai15,pal19,cap20taj}.

Another common theme has been a trade-off between interpretability of model parameters and model complexity. In order to make the CP and the BDSP amenable to infer relevant quantities such as prevelance, one needs to both impose more modeling assumptions and incorporate more data. Recall that under complex epidemiological models, the model becomes unidentifiable unless we pre-specify some of the parameter values or incorporate independent sources of information. However, incorporating other sources of information and their corresponding sampling models can also create biases when the models are misspecified. We envision future research that incorporates robust models against model  misspecification such as the adaptive preferential sampling. As it is common in many other areas of science, the strive for a balance between realism on one side, and simplicity and interpretability on the other, is going to be an essential focus of future work.

Our discussion has omitted other phylodynamic models that have been used to track the evolution of the pandemic such as phylogeography models \citep{lemey2020accommodating}, structured coalescent models \citep{Mueller2017}, coalescent with recombination \citep{muller2021recombination} and models of within-host-evolution \citep{Jones2018}. We have also omitted model selection from our discussion and refer the reader to \cite{lewis2014posterior}. We do not discuss other data-quality associated statistical challenges such as sequencing errors \citep{tur20,mor21} and underreporting of case count data \citep{wu20}.

\section{Acknowledgments}
J.A.P. acknowledges support from National
Institutes of Health Grant R01-GM-131404.

\vspace{2cm}

\begin{appendix}
\section{Data analyses}

\subsection{Phylodynamic analysis in California}

Case counts for Figure~\ref{fig:cali} panel (A) were obtained from the New York Times repository (\url{ https://github.com/nytimes/covid-19-data}). 

\noindent Molecular sequences for Figure~\ref{fig:cali} panel (B) were obtained from the GISAID repository (accession codes of the sequences used can be retrieved at\\ \url{https://github.com/JuliaPalacios/phylodyn/blob/master/data/California\_statscience\_ack.txt}). 
Given the molecular sequences, a viral phylogeny was obtained from a genetic distance-based method called serial UPGMA \citep{dru00}. Conditionally on the viral phylogeny, the EPS posterior was inferred with a Bayesian nonparametric method described in \cite{pal12}. The posterior approximation is based on INLA.

\subsection{Analysis of SARS-CoV-2 sequences from Washington State}

The two sets of 100 molecular sequences were analyzed independently with BEAST 2 with the same model and MCMC parameters. We run the chains for $20 \times 10^6$ iterations, thinning every $1000$ and with a burnin of $10 \times 10^6$ iterations. We used the Extended Bayesian Skyline prior on $N_e(t)$ \citep{hel08}, the HKY mutation model with empirically estimated base frequencies \citep{hky}, and the mutation rate fixed to $9 \times 10^{-4}$ substitutions per site per year. The two phylogenies obtained are the maximum clade credibility trees of the posterior distributions obtained with TreeAnnotator \citep{bou19}. Accession codes of the sequences can be retrieved at 
\url{https://github.com/JuliaPalacios/phylodyn/blob/master/data/Washington\_statscience\_ack.txt}. Details on the analyses done are included in the main text. 

\subsection{MDS analysis}
Molecular sequences used to generate Figure~\ref{fig:mds} were obtained from GISAID. We analyzed 9 samples of 100 sequences (3 samples of 100 sequences per country) independently with BEAST2 \citep{bou19} with the same model and MCMC parameters. We run the chains for $50 \times 10^{6}$ iterations. We used the Bayesian Skyline prior on $N_{e}(t)$, the HKY mutation model, and the mutation rate fixed to $9 \times 10^{-4}$ substitutions per site per year. Posterior samples were thinned to 100 samples. Pairwise distances were obtained using \citet{Kim2020}. Accession codes of the sequences can be retrieved at\\ \url{https://github.com/JuliaPalacios/phylodyn/blob/master/data/CanSweUk\_statscience\_ack.txt}

\section{Glossary of terms}

\noindent\textbf{Anthropogenic selection.} A process where human-induced environmental changes, such as use of antiviral drugs, alter the direction and magnitude of selection.

\noindent\textbf{Antigenic evolution} An evolutionary response of pathogens to host's antibody-mediated immunity selective pressures.  

\noindent\textbf{Consensus sequence.} A consensus individual's sequence consists of those nucleotides with highest frequency at each position in an alignment (aligned to a reference genome) of multiple reads \citep{grubaugh2019amplicon}; usually only those nucleotides with high frequency and high coverage (multiple reads per nucleotide) are used in the analyses.

\noindent\textbf{Endemic equilibrium} A state at which the disease dynamics is in a steady state so the disease persists in the population. 

\noindent\textbf{Locus.} The physical location of a specific gene on a chromosome. Here we assume there is no recombination within the locus.

\noindent\textbf{Mutation.} An alteration in a genetic sequence such as substitution, insertions, deletions, etc.

\noindent\textbf{Recombination.} The exchange of genetic material between parental genomes by the breakage and rejoining of chromosomes, producing offspring genomes that carry genetic information distinct from its parental genomes. 

\noindent\textbf{Selection.} A non-random difference in reproduction among individuals, often due to differential survival to specific environments, ensuring the transmission of beneficial traits to succeeding generations.

\noindent\textbf{Substitutions.} A type of mutation where a single nucleotide (``chemical letter'') is replaced with a different nucleotide.


\end{appendix}
\bibliographystyle{imsart-nameyear} 
\bibliography{ref.bib}       


\end{document}